\documentstyle[epsf,psfig]{mn}

\begin{document}
 \title[LMC sources]
     {Optical studies of two LMC X-ray transients: RX J0544.1--7100
 and RX J0520.5-6932
  \thanks{Partially based on observations collected at the South African Astronomical
 Observatory and the European Southern Observatory, Chile (ESO N64.H-0059)}}

 \author[M.J.Coe et al.]
 {M.J. Coe$^{1}$, I. Negueruela$^{2,3}$, D.A.H. Buckley $^{4}$, 
 N.J. Haigh$^{1}$
\& S.G.T.Laycock$^{1}$\\
 $^{1}$Physics and Astronomy Dept., The University, Southampton, SO17
 1BJ, UK. \\
 $^{2}$SAX SDC, Agenzia Spaziale Italiana, c/o Telespazio, via Corcolle
 19, 00131 Roma, Italy.\\
 $^{3}$Observatoire de Strasbourg, 11 rue de l'Universite, Strasbourg
 67000, France.\\
 $^{4}$South African Astronomical Observatory, PO Box 9, Observatory
 7935, South Africa.
 }

 \date{Accepted \\
 Received : Version November 2000\\
 In original form ..}

 \maketitle

 \begin{abstract} 

 We report observations which confirm the identities of the optical
 counterpart to the transient
 sources RX J0544.1-7100 and RX J0520.5-6932.  The counterparts are
 suggested to be a B-type stars.  Optical data from the observations
 carried out at ESO and SAAO, together with results from the OGLE data
 base, are presented. In addition, X-ray data from the RXTE all-sky
 monitor are investigated for long term periodicities. A strong
 suggestion for a binary period of 24.4d is seen in RX J0520.5-6932
 from the OGLE data.

 \end{abstract}

  \begin{keywords}
 stars: emission-line, Be - star: binaries - infrared: stars - X-rays: stars -
 stars: pulsars
  \end{keywords}

 \section{Introduction}

 The Be/X-ray and supergiant binary systems comprise the class of
 massive X-ray binaries.  A survey of the literature reveals that of
 the 96 proposed massive X-ray binary pulsar systems, 67\% of the
 identified systems fall within the Be/X-ray group of binaries.  The
 orbit of the Be star and the compact object, a neutron star, is
 generally wide and eccentric.  The optical star exhibits H$\alpha$
 line emission and continuum free-free emission (revealed as excess
 flux in the IR) from a disk of circumstellar gas. Most of the Be/X-ray
 sources are also very transient in the emission of X-rays.


 \begin{figure}
 \begin{center}
 \psfig{file=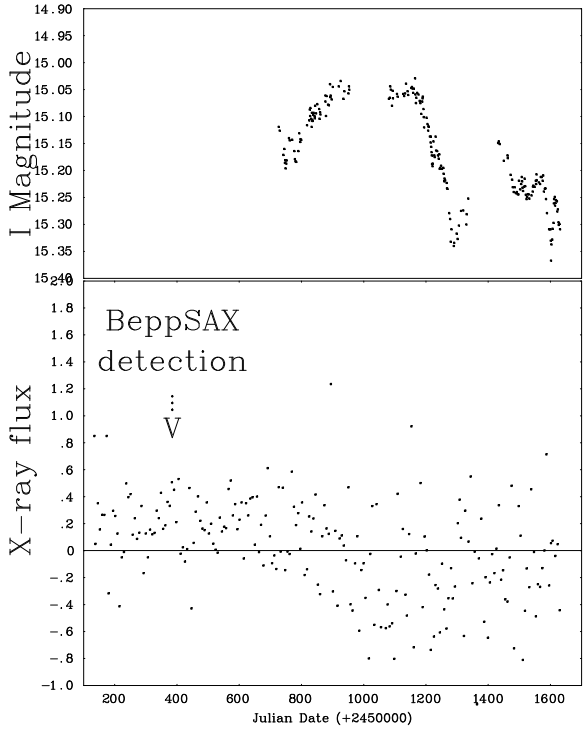,width=3in}
 \end{center}
 \caption{Top panel : the optical lightcurve of RX J0544.1--7100
 obtained from the
 OGLE data. Lower panel : the X-ray lightcurve from the RXTE ASM averaged over 7
 day intervals.}
 \end{figure}


 \subsection{RX J0544.1--7100}

 The source 1SAX J0544.1--710 was detected by BeppoSAX in October 1996
 (Cusumano et al, 1998) at a flux corresponding to a luminosity of 9 x
 $10^{35}$ erg/s.  Their observations revealed an X-ray pulse period
 of 96s. The transient, variable nature of the source was reported by
 Haberl et al (1998) when they linked the BeppoSAX source with a ROSAT
 source, RX J0544.1--7100 (Haberl \& Pietsch 1999).  Haberl et al also
 reported a possible optical counterpart lying within the 8 arcsec
 ROSAT error circle. Subsequently, Haberl \& Pietsch (1999) reported
 more details of the ROSAT source and refined the error circle to just
 3.3 arcsec radius which includes one obvious optical counterpart.
 The PSPC spectrum indicated that it was the hardest source in their
 sample of 27 LMC objects they studied. The totality of the X-ray
 behaviour strongly suggests that the object is a member of the
 Be/X-ray binary class.

 Reported here are optical, infra-red and X-ray measurements of the system. 
 The data confirm the proposed identity of the
 counterpart to RX J0544.1--7100, and the counterpart is shown to be most
 consistent with a main sequence B0V star. 

 \subsection{RX J0520.5-6932}

 The source RX J0520.5--6932 was discovered by ROSAT (Schmidtke et al,
 1994) at a luminosity of 5 x $10^{34}$ erg/s and identified with a
 V$\sim$14 magnitude star.  Optical spectral observations carried out
 by Schmidtke et al indicated a O8e type star with radial velocity
 measurements consistent with LMC membership. Since the source was not
 detected by Einstein it is probably exhibiting X-ray variability
 consistent with it being a member of the Be/X-ray binary group of
 sources - though its spectral classification is unusually early
 (Negueruela 1998). Further optical observations by Schmidtke et al
 (1996) revealed evidence for small photometric changes up to a few
 tenths of a magnitude.

 \section{OGLE data}

 The Optical Gravitational Lensing Experiment (OGLE) is a long term
 observational program with the main goal of searching for dark, unseen
 matter using the microlensing phenomenon (Udalski et al. 1992). 
 

 In general the OGLE data cover the period June 1997 to February 2000
 and primarily consist of I band observations, though some observations
 were also taken in the V band. The optical counterparts 
were identified in the OGLE data base and all the photometric
 measurements extracted. Figure 1 (top panel) shows the optical
 lightcurve for RX J0544.1-7100 obtained from the OGLE I band data.

 The OGLE data of RX J0520.5-6932 are shown in Figure 2.

 \begin{figure}
 \begin{center}
 \psfig{file=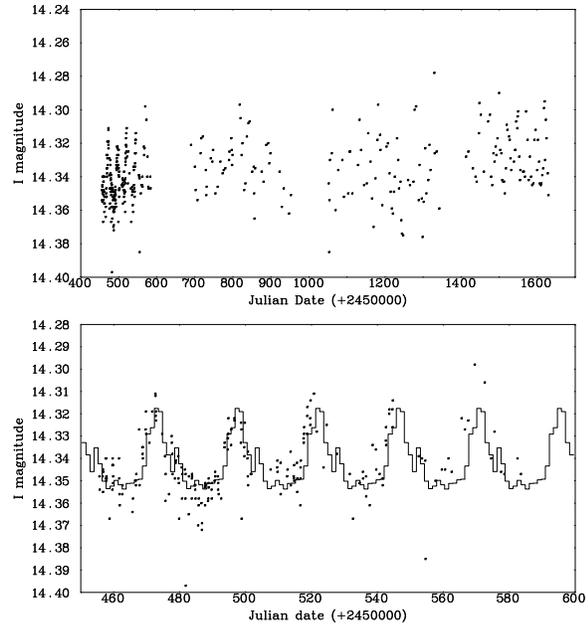,width=3in}
 \end{center}
 \caption{Upper panel: the optical lightcurve of RX J0520.5--6932
 obtained from the
 OGLE data. Lower panel: comparison of the first season of 
 OGLE data with the average
 modulation template.}
 \end{figure}

 Figure 3 shows the power spectrum of RX J0544.1-7100 obtained using
 the Lomb-Scargle technique 
 on 219 I band data points obtained over the period 5 Oct 1997 -
 27 March 2000. Periods in the range 10 - 500 days were
 investigated. The largest peak shown in Figure 3 corresponds to a
 period of 286d and undoubtably arises from the variability on several
 long timescales evident in Figure 1.

 \begin{figure}
  \psfig{file=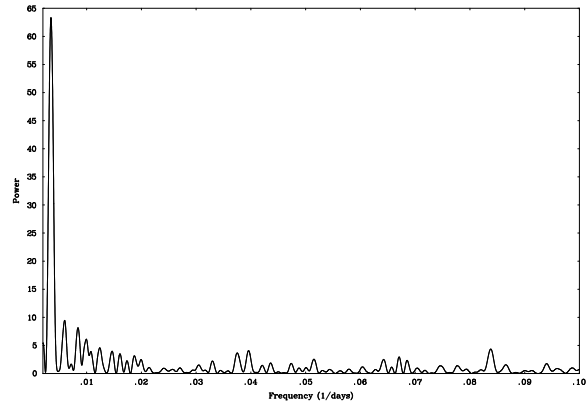,angle=-90,width=3in}
 \caption{The power spectrum obtained using the Lomb-Scargle algorithm on the
 OGLE data of RX J0544.1--7100. The highest peak corresponds to a
 period of 286d.}
 \end{figure}

 Figure 4 shows the Lomb-Scargle power spectrum obtained for RX
 J0520.5-6932 using the same search parameters as those used for RX
 J0544.1--7100. In this case there is a very clear and strong peak
 corresponding to a period of 24.45d.  The lower panel in Figure 2
 shows the average folded lightcurve compared to the first, and most
 comprehensive, set of OGLE data.
 Phase zero has been set to
 JD2450008.6 corresponding to the peak of the optical flux.

\begin{figure}
\psfig{file=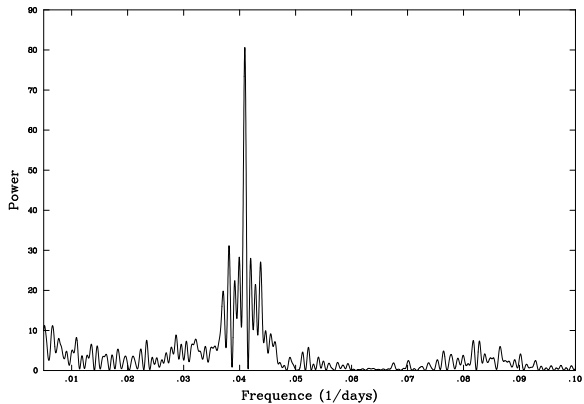,angle=-90,width=3in}
 \caption{The power spectrum obtained using the Lomb-Scargle algorithm on the
 OGLE data of RX J0520.5--6932. The highest peak corresponds to a
 period of 24.45d.}
 \end{figure}

 \section{Rossi X-ray Timing Explorer all-sky monitor data}

 Data from the RXTE all sky monitor experiment were obtained from the
 public archive for RX J0544.1--7100. 
 Since the signal from this object is very weak, the
 data shown in the lower panel of Figure 1 have been averaged over 1
 week intervals.

 Looking carefully at the X-ray lightcurve in Figure 1 it is clear that
 a significant X-ray signal is only present in the first half of the
 data run. Therefore only X-ray data covering the epoch TJD 100-1000 were
 searched for periodic signals.  The raw daily data for this
 period were then analysed using the same Lomb-Scargle 
 algorithmic technique
 as the optical data. No periodic behaviour was identified in these
 X-ray data.

 The peak ASM X-ray luminosity may be determined from Figure 1 as reaching
 values of $\sim$0.5 cts/s. This corresponds to a 
peak luminosities of $\sim$6 x $10^{35}$
 erg/s and is in good agreement with the BeppoSAX value of 8 x
 $10^{35}$ erg/s quoted earlier in this paper.

 \section{Optical and IR photometry}

 The sources were observed by the SAAO 1.0m telescope in 1996, 1999 and
 2000. The exact dates of the individual observations are given in
 Tables 1 and 2. The data were collected using the Tek8 CCD giving a field of
 approximately 3 arcmin and a pixel scale of 0.3 arcsec per pixel.
 Observations were made through standard Johnson UBVRI and
 Str{\"o}mgren-Crawford uvby$\beta$ filters.  The data were reduced
 using IRAF and Starlink software, and the instrumental magnitudes were
 corrected to the standard system using E region standards.

 In addition, IR data on RX J0544.1-7100 were obtained from the public
 archives of the 2MASS survey of the LMC. Though the data were taken
 about a year before our optical observations, they are included here
 so that possible IR emission from the circumstellar disk could be
 investigated. There are no reported IR data on RX J0520.5-6932 in
 the same catalogue.

 All the resulting photometric magnitudes are given in Tables 1 and 2.

 \begin{table}
  \centering
  \caption{Optical and IR photometry of the counterpart to RX
 J0544.1--7100.}
  \begin{tabular}{lccc}

 Band & 21 Mar 1998&7 Jan 1999&OGLE\\
 &(2MASS data)&&average\\
 &&&\\
 B&& 15.39$\pm$0.05&\\
 V&& 15.36$\pm$0.05&15.27\\
 R&& 15.26$\pm$0.05&\\
 I&&&15.14\\
 u&& 15.63$\pm$0.05&\\
 v&& 15.45$\pm$0.05&\\
 b&& 15.45$\pm$0.05&\\
 y&& 15.30$\pm$0.05&\\
 J&15.18$\pm$0.06&&\\
 H&14.97$\pm$0.09&&\\
 K&15.19$\pm$0.19&&\\

  \end{tabular} 
 \end{table}

 \begin{table}
  \centering
  \caption{Optical and IR photometry of the counterpart to RX
 J0520.5-6932.}
  \begin{tabular}{lccc}

 Band & 4 Oct 1996&17 Jan 2000&OGLE\\
 &&&average\\
 &&&\\
 U&14.1$\pm$0.1&&\\
 B& 14.44$\pm$0.01&&14.39\\
 V& 14.43$\pm$0.01&&14.45\\
 R& 14.36$\pm$0.03&&\\
 I&&&14.34\\
 u&& 14.4$\pm$0.1&\\
 v&& 14.55$\pm$0.03&\\
 b&& 14.43$\pm$0.02&\\
 y&& 14.33$\pm$0.02&\\

  \end{tabular} 
 \end{table}

 The average OGLE values over all their data come from Udalski (private
 communication). Of course, it is obvious from Figures 1 and 2 that the
 sources have not been constant and so we should not expect perfect
 agreement between the OGLE average values and the specific values
 reported here.

 \section{Optical spectroscopy}

 A red spectrum of RX J0544.1-7100 is shown in Figure 5, was obtained
 from the 1.9m SAAO observatory on 9 January 1999 using the Cassegrain
 spectrograph with the SITe2 CCD detector.  Though of a low
 signal-to-noise ratio, the spectrum, clearly shows H$\alpha$ in
 emission, though little can be determined about the line shape.  The
 H$\alpha$ line has an equivalent width of EW = -7$\pm$1\AA ~and a
 central position of 6568$\pm$1\AA.

 \begin{figure}
\psfig{file=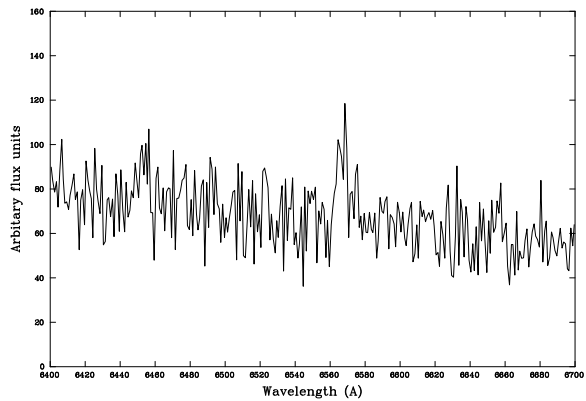,angle=-90,width=3in}
 \caption{Red optical spectra obtained of the candidate to RX
 J0544.1--7100.}
 \end{figure}

 Blue spectroscopy of both sources were obtained on 1st November 1999 using 
 the ESO 1.52-m telescope at La Silla Observatory, Chile. The telescope was 
 equipped with the Boller \& Chivens spectrograph + \#32 holographic 
 grating. The nominal dispersion for this configuration is $\sim 
 0.5$~\AA/pixel, while measurements of the FWHM of arc lines indicate a 
 spectral resolution of $\approx 1.4$~\AA\ at $\sim \lambda4500$~\AA.
 The blue spectra are shown in Figures 6 and 7. For RX J0544.1--7100 
 the H$\beta$ line has an
 equivalent width of EW = -0.20$\pm$0.02\AA.

 \section{Discussion}

 \subsection{RX J0544.1--7100}

 \subsubsection{Spectral Class and LMC membership}

 The spectrum of the optical counterpart to RX~J0544.1-7100 in the
 classification region is displayed in Figure 6, together with that of
 the B0V standard $\nu$ Ori. The spectrum of a Be star of similar
 spectral class is included for comparison. This is HD~161103, given as
 B0.5III-Ve by Steele et al. (1999). 


Though the
spectrum of RX~J0544.$-$700 has a low signal-to-noise ratio, photospheric
lines are relatively clear and strong shortwards of $\sim 4200$~\AA.
Weak Si{\sc
 iv}~$\lambda4089$\AA\ and He{\sc ii}~$\lambda4686$\AA\ indicate that
 the spectral type of RX~J0544.1-7100, as in most counterparts to
 Be/X-ray binaries, is close to B0. Since this lines are weak and
 He,{\sc ii}~$\lambda4200$\AA\ is not present, the star cannot be
 earlier.  The weakness of the Si{\sc iii} triplet and C{\sc iii} shows
 that it is not an evolved star. Since the presence of He{\sc ii} lines
 indicates a spectral type earlier than B0.5, we classify the optical
 counterpart to RX~J0544.1-7100 as B0V, with an uncertainty of half a
 spectral subtype (due to the low SNR).

 If we assume this spectral class of B0V then we can determine the line
 of sight extinction to our source from the photometry in Table 1. A
 star of this spectral type has absolute magnitudes of B=--4.3, V=--4.0
 and R=--3.87. The distance modulus to the LMC is 18.5$\pm$0.2
 magnitudes (Westerlund 1997). In order to get agreement between a
 reddened version of these photometric magnitudes and our observations
 it is necessary to assume E(B-V)=0.26$\pm$0.06. If we also consider
the uncertainty of a half sub-class in the spectral type, then this
uncertainty could be as large as 0.08. Schwering \&
 Israel (1991) quote a foreground reddening to the LMC in the range
 0.07 to 0.17 and from their Figure 7b it can be seen that RX
 J0544.1-7100 lies in the region of highest reddening
 where E(B-V)$\sim$0.15. Our result is consistent within errors 
with their value.
If the reddening to this system proves to be higher than expected then 
some of this may be accounted
 for from local circumstellar extinction around the Be star. Using the
 H$\alpha$ EW = -7\AA ~and Equation 5 from Fabregat \& Torrejon (1998),
 we can estimate that $E^{cs}$(B-V)=0.03. 

 Interestingly, the IR photometry presented here from the 2MASS survey
 indicate more than just a high level of interstellar reddening. The
 intrinsic (J--K) for a B0V star is -0.23, where as we have an observed
 value of -0.01. Using our value of E(B--V)=0.26 gives E(J--K)=0.14,
 and hence a predicted observed value of (J--K)=-0.09. Though the errors
 on the 2MASS values, especially K, are rather large, there is a suggestion
 of further reddening in the IR band arising from the presence of a
 circumstellar disk.

 On the question of LMC membership 
 there can be little doubt that this system lies in the LMC.  The
 position of the H$\alpha$ emission line corresponds to a red shift of
 228$\pm$45 km/s, and the redshift of the absorption lines in the blue
 spectrum to an average value of 220$\pm$80 km/s. Both of these data
 sets are consistent with LMC membership (for example, Fischer, Welch
 \& Mateo (1993) quote a mean velocity for a cluster in the LMC of
 251$\pm$2 km/s).

 \subsubsection{X-ray luminosity}

 \subsubsection{Optical variability}

 It is almost certain that the largest peak in the OGLE power spectrum
 (at 286d) reflects the global changes taking place in the
 luminosity. The total data run is only approximately 900d, so it is
 not really long enough to claim that a $\sim$300d period could be a
 coherent modulation in any convincing manner. It is much more likely
 that the fluctuations seen are similar to those seen in other Be
 stars. Certainly the two relatively deep minima at TJD $\sim$1290 and
 $\sim$1600 will be providing a lot of the power seen around
 $\sim$300d. 

 \subsection{RX J0520.5-6932}

 \subsubsection{Binary period}

 The only detection of this source in X-ray has been reported by
 Schmidtke et al (1994). Their ROSAT observations were carried out on
 11 Feb 1991 which, interestingly, correspond to a binary phase of 0.90
 (phase 0.0 has been defined here as the peak optical flux). In
 addition, Schmidtke et al (1999) report a non-detection by ROSAT at
 binary phase 0.58 (22 July 1995). Consequently it is possible that the
 X-ray flux is modulated in the same manner as the optical, though the
 evidence so far is rather sparse.

 In the case of RX J0520.5-6932 the optical modulation shown in Figure
 2 bears a striking resemblance in shape to that seen in A 0538-66
 (Densham et al, 1983), but at a much smaller magnitude. In that source
 the optical outburst reached a value of 2.5 magnitudes above the
 baseline whereas we only see variations of 0.03
 magnitudes. (More recent work by Alcock etal, 2000 has shown that the
 optical outburst size in A 0538-66 has decreased to 0.4-0.6
 magnitudes). Nonetheless the periods are not very different (16.6d in A
 0538-66 and 24.4d in our object) and the optical counterparts are both
 Be stars. Charles et al  (1983) attribute the optical outbursts in A
 0538-66 to localised Roche lobe overflow induced by the close passage
 of the neutron star.

 \subsubsection{Spectral classification and LMC membership}

 The spectrum of the optical counterpart to RX~J0520.5.--6932 in the
 classification region is displayed in Figure 7, together with that of
 two standard stars. 

As noted by Schmidtke et al.
(1994), the photospheric lines are extremely shallow. This is, at least in
part, due to the obvious presence of emission components affecting all
H\,{\sc i} and He\,{\sc i} lines. H$\beta$ is in emission with an EW of
-0.5\AA, while H$\gamma$ is completely filled in. It must be pointed out,
however, that the He\,{\sc ii} and metallic lines where no infilling
is expected are also very shallow. 

The presence of relatively strong He\,{\sc ii} lines clearly
identifies the object as an O-type star. In Fig 8, the MK standard
stars AE Aur (O9.5V) and 10 Lac (O9V) are also shown as
comparison. The well-marked Si\,{\sc iv} lines in the wings of
H$\delta$ indicate a spectral type close to O9, while their moderate
strength and the absence of obvious Si\,{\sc iii} or O\,{\sc ii} lines
indicates a main-sequence star. The presence of C\,{\sc iii} $\lambda$
4650\AA\ and the strength of the Si\,{\sc iv} lines makes the O8V
spectral type proposed by Schmidtke et al. (1994) seem a bit too
early. However, the shallowness of the lines makes the spectral
classification slightly insecure. We will therefore adopt a spectral
type O9V, allowing at the most one spectral subtype uncertainty.

Such a classification is consistent with our photometric measurements
and an E(B-V)=0.32 (very similar to the value obtained for RX
J0544.1-7100).  

Finally from the position of the H$\beta$ line we obtain a red shift
of 345$\pm$30 km/s which, though a slightly high value, is consistent
with LMC membership.

\subsection{General remarks}

Since the system parameters of these objects seem to indicate a
Be/X-ray binary system then the Corbet diagram (Corbet et al, 1999)
may be used to estimate the missing timing parameters. For RX
J0544.1--7100 we have $P_{pulse}$=96s and hence a likely $P_{orb}$
would be just over 100d similar to the well-studied system A0535+26
($P_{pulse}$=104s, $P_{orb}$=111d). In the case of RX J0520.5--6932 we
have $P_{orb}$=24.45d and hence a probable $P_{pulse}$=3-4s, making it
similar to 4U 0115+63 ($P_{pulse}$=3.6s, $P_{orb}$=24.3d).

\begin{figure*}
\psfig{file=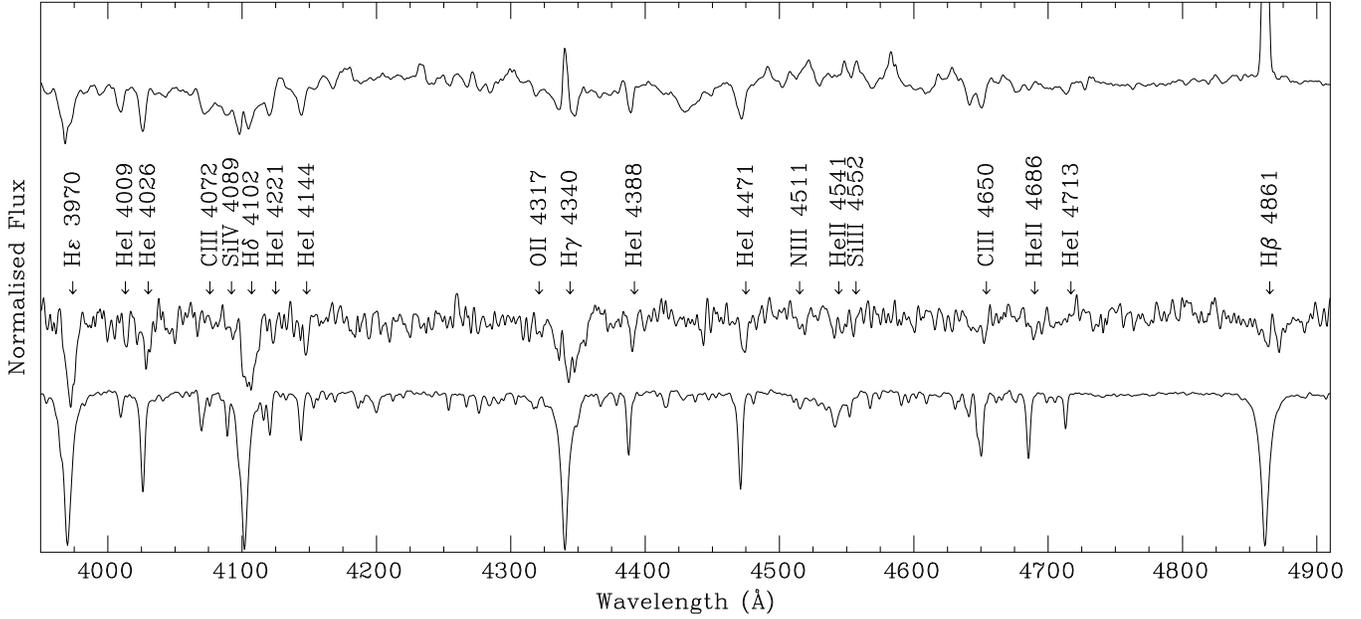,angle=-90,width=7in}
\caption{The blue spectrum of the optical counterpart to RX~J0544.1-7100
(middle) compared to that of the B0V standard $\nu$ Ori (bottom) and the
B0.5e star HD~161103 (top). All spectra have been normalised
by division into a spline fit to the continuum and smoothed with a Gaussian
filter ($\sigma$ = 0.8~\AA).}
\end{figure*}

\begin{figure*}
\psfig{file=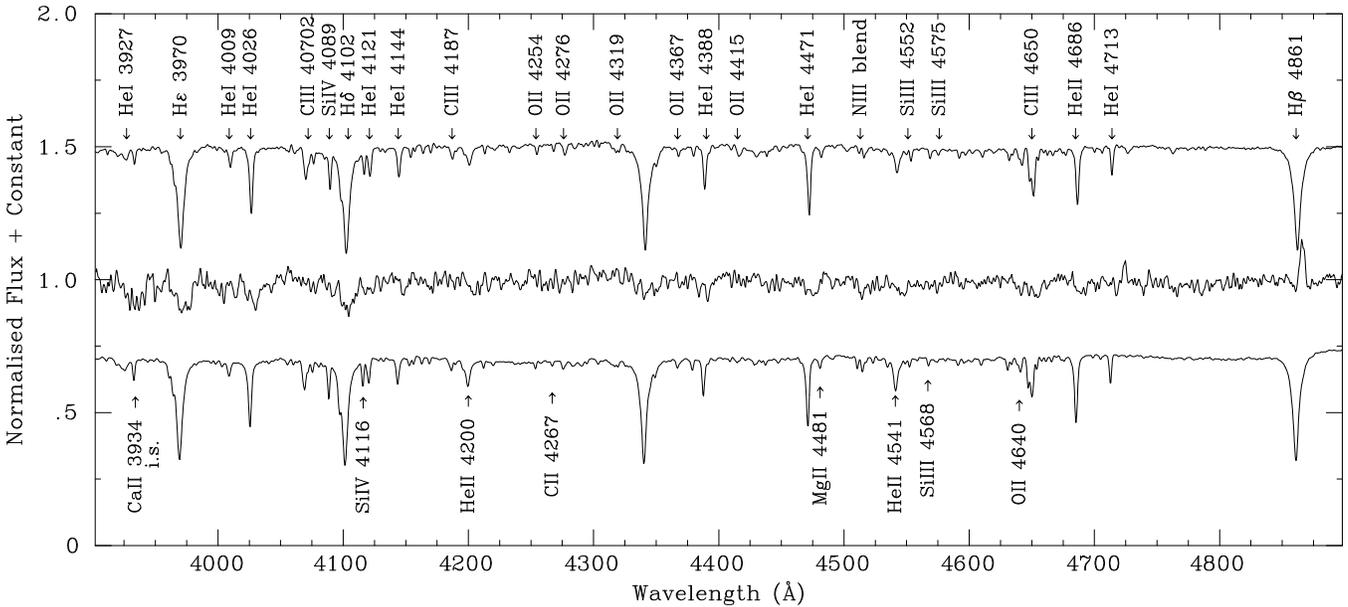,angle=-90,width=7in}
\caption{The blue spectrum of the optical counterpart to 
RX~J020.5--6932 (middle) compared to that of the 09.5V standard AE Aur
(upper) and the O9V standard 10 Lac (lower). All spectra have been normalised
by division into a spline fit to the continuum and smoothed with a Gaussian
filter ($\sigma$ = 0.6~\AA).} 
\end{figure*}



\section*{Acknowledgments}

We are extremely grateful to Andrzej Udalski and the OGLE team for
providing us with their data.  We are also grateful to the very
helpful staff at the SAAO for their support during these observations.
All of the data reduction was carried out on the Southampton Starlink
node which is funded by the PPARC.  NJH and SGTL are in receipt of a
PPARC studentships. During part of this work 
IN was supported by an ESA external research
fellowship. 

This publication makes use of data products from the Two Micron All
Sky Survey, which is a joint project of the University of
Massachusetts and the Infrared Processing and Analysis
Center/California Institute of Technology, funded by the National
Aeronautics and Space Administration and the National Science
Foundation.

\bsp

\end{document}